\documentclass[preprint,12pt]{elsarticle}

\usepackage{graphics}

\usepackage{amssymb}
\usepackage{amsmath}

\journal{NIM A}

\begin{document}

\begin{frontmatter}


\title{Search for Dark Matter in the Sun with the ANTARES Neutrino Telescope in the CMSSM and mUED frameworks}

\author{J. D. Zornoza on behalf of the ANTARES collaboration}

\address{IFIC - Instituto de F\'isica Corpuscular, Ed. Instit. de
  Investigaci\'on, AC 22085, E-46015, Valencia, Spain}

\begin{abstract}
ANTARES is the first neutrino telescope in the sea. It consists of a
three-dimensional array of 885 photomultipliers to collect the
Cherenkov light induced by relativistic muons produced in CC
interactions of high energy neutrinos. One of the main scientific
goals of the experiment is the search for dark matter. We present here
the analysis of data taken during 2007 and 2008 to look for a WIMP
signal in the Sun. WIMPs are one of the most popular scenarios to
explain the dark matter content of the Universe. They would acumulate
in massive objects like the Sun or the Galactic Center and their
self-annihilation would produce (directly or indirectly) high energy
neutrinos detectable by neutrino telescopes. Contrary to other
indirect searches (like with gamma rays or positrons), the search for
neutrinos in the Sun is free from other astrophysical contributions,
so the interpretation of a potential signal in terms of dark matter is
much more robust.

\end{abstract}


\end{frontmatter}



\section{Introduction}
\label{introduction}

One of the most relevant questions in physics is the nature of dark
matter. Since the first hints about this issue were pointed out by
Zwicky in 1933~\cite{zwicky}, a rich set of experimental proofs have
been obtained to confirm the fact that most of the matter in the
Universe is non-luminous. These proofs include the observations of
WMAP~\cite{wmap}, the results on the Big Bang
Nucleosynthesis~\cite{jedamzik}, the rotation curves of
galaxies~\cite{rubin} and the studies of highly
red-shifted Ia supernovae~\cite{kowalski}. The conclusion is that our
Universe is made of 73\% of dark energy and 27\% of matter. Moreover,
most of the matter (about 80\%) is non-baryonic, whose nature is
unknown. We know that this non-barionic component cannot be dominated
by neutrinos, since they travel at relativistic velocities and
therefore they cannot explain the structure of the Universe observed
today. Consequently, the candidates for explaining most of dark matter
are not in the Standard Model. There are several models to explain
such particles. One of the most accepted candidates come from
Weakly-Interacting Massive Particles (WIMPs), following the
requirements of having an interaction cross-section of the order of
that for weak interactions (to reproduce the right relic density)
and being massive (to account of the gravitational effects). They also
have to be stable. Among the frameworks which are able to provide such
particles we will study here two of the most popular ones:
SuperSymmetry (SUSY) and Universal Extra Dimensions (UED). In
particular, we have considered two constrained versions: CMSSM~\cite{cmssm} and
mUED~\cite{mued}. The candidate particles in each case are the neutralino and the
lightest Kaluza-Klein particle. Their stability is explained by the
conservation of the R-parity in the case of SUSY and KK-parity in the
case of UED.

The potential detection of WIMPs in neutrino telescopes is based on the
hypothesis that they would accumulate in massive objects like the Sun,
the Galactic Center or the Earth due to the loss of energy through
elastic scattering. If they are Majorana particles they
would self-annihilate and produce secondaries which in turn would
yield high energy neutrinos detectable by neutrino telescopes. For the
case of Kaluza-Klein particles, direct production of neutrinos is also
possible. In the analysis presented here we focus on the Sun. This is
very advantageous since a potential detection of high energy neutrinos
from the Sun direction would be a very clean signal of dark
matter. In constrast to other indirect searches, there would not be other
astrophysical explantions, as it happens for instance, when an excess
of gammas or positrons is detected, for which other astrophysical
scenarios have to be taken into account.

The analysis presented here is an binned search in the data taken by
the ANTARES neutrino telescope during 2007 and 2008.

\section{The ANTARES neutrino telescope}
\label{antares}

ANTARES~\cite{antares} is the largest neutrino telescope in the
Northern hemisphere. It is located at (42$^{\circ}$ 48' N, 6$^{\circ}$
10' E) at a depth of 2475~m in the Mediterranean Sea, 42~km off-shore
Toulon, France. It consists of 885 10-inch photomultipliers (PMTs)
distributed over 12 vertical lines together forming a
three-dimensional array. Each PMT is housed in a pressure-resistant
glass sphere (Optical Module, OM), together with a gel for optical
coupling and a mu-metal cage to attenuate the Earth's magnetic
field. The OMs are arranged in triplets (storeys) for a better
rejection of the optical background. There are 25 storeys per line
except for one of the lines which has only 20 storeys, since the upper
part of this line contains devices for acoustic detection tests. The
PMTs are pointing 45$^{\circ}$ downward for a better detection of
upgoing events. The length of the lines is 450~m and the horizontal
distance between neighbouring lines is 60-75~m. There is also an
additional line with devices aimed to monitor the environment (sea
current velocity, pressure, temperature, salinity, etc.)

The detection principle is as follows. When a high energy neutrino
interacts via a charged current (CC) interaction close or inside the
detector (in the water or in the rock below), a relativistic muon is
produced. This muon induces Cherenkov light when travelling through
the water, which is detected by the photomultipliers. The information
of the time~\cite{timing} and position is used in order to reconstruct
the direction of the muon and therefore the direction of the
neutrino\footnote{ There are other possible signatures, not used in
  this analysis: CC interactions of electron and tau neutrinos as well
  as the NC interactions of all flavours, produce shower events,
  which, given the granularitiy of the detector, are seen as point
  sources of light.}. The track of the muon is reconstructed with a
robust fitting algorithm based on a $\chi^{2}$ test which uses the position and time of the
hits~\cite{bbfit}. It gives an angular resolution of $\sim$2 degrees at
energies of tens of GeV. The $\chi^{2}$ provided by the fit algorithm
is used for the selection of well reconstructed tracks, as explained in
Section~\ref{cuts}.

\section{Data and Monte Carlo simulation}
\label{simulation}

\begin{figure}
\includegraphics[width=\linewidth]{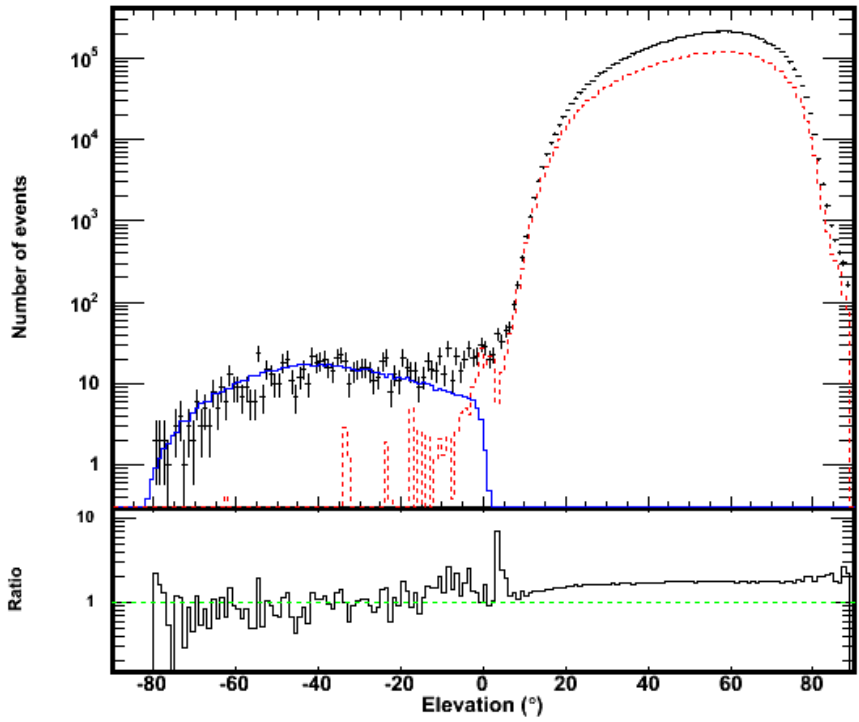} \\
\caption{Comparison between data and simulation: elevation of
  events. The distributions show the simulated atmospheric muons
  (red), the simulatd atmospheric upgoing-neutrinos (blue), and the
  reconstructed data (black). The ratio of data over simulation is
  shown below the corresponding plot.}
\label{mcdata}
\end{figure}

In this analysis we have used data taken in 2007 and
2008. During 2007, there were only 5 lines connected. In May 2008 the
detector was completed. There were three different
configurations: 9, 10 and 12 lines (since May 2008). The number of
active days was 185.5 and 189.8 for 2007 and 2008, respectively. The
un-filtered data are dominated by the background from to atmospheric
muons, produced by the interaction of cosmic rays in the
atmosphere. Most of this background is avoided by accepting only
up-going events (since muons cannot traverse the Earth) but a fraction
of them are mis-reconstructed as up going. The other source of
background is due to atmospheric neutrinos, i.e. neutrinos produced in
the interactions of cosmic rays in the atmosphere. In this case, the
neutrinos can traverse the Earth, so it is an irreducible background
which is discriminated using the fact that this is a diffuse
flux, while the signal we are looking for is peaked in the Sun's
direction. In Fig.~\ref{mcdata} we show a comparison between data
and simulation of these two types of backgrounds. It is important to
note that the estimation of the background used in the analysis comes
directly from scrambled data, in order to reduce the effect of
systematic uncertainties associated with the efficiency of the detector
and the assumed flux.

In order to simulate the signal, the WimpSim package is
used~\cite{wimpsim}. This program allows for a computation of the
neutrino flux arriving at the Earth for different annhiliation
channels. The main physical processes are taken into account
(interactions in the Sun, regeneration of tau leptons, oscillations in
the propagation, etc.) Once the neutrino flux is computed for the
channels we are interested in, we can weight the events according to the
branching ratios of the model under study. As already mentioned, the
two frameworks we have considered here are CMSSM and mUED. The main
channels for our detector are $W^+W^-$, $b\bar{b}$ and
$\tau\bar{\tau}$ for CMSSM and $c\bar{c}$, $b\bar{b}$,
$\tau\bar{\tau}$, $t\bar{t}$ and $\nu\bar{\nu}$ for mUED. In addition
to the incoming flux, we have to estimate, through simulation, the
response of the detector, i.e. the effective area, which relates this
flux with the number of detected events for a given set of cuts.

\section{Cut optimization}
\label{cuts}

\begin{figure}
\includegraphics[width=\linewidth]{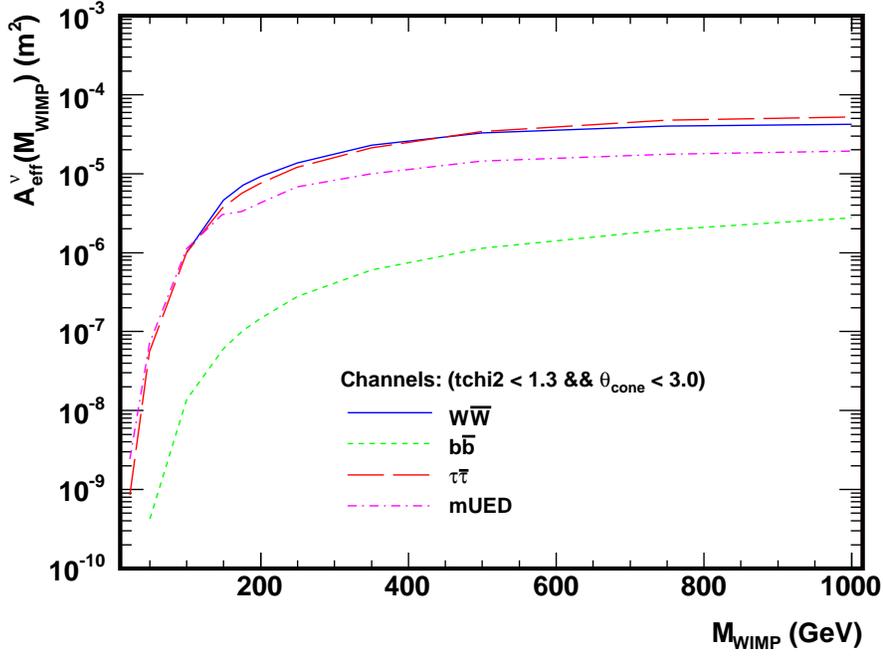}
\caption{Normalized effective area for different WIMP masses.}
\label{aeff_channels}
\end{figure}

As mentioned above, we have followed a binned search strategy, i.e. we
compare the number of events observed in a cone around the Sun
direction with the expected background, looking for an excess which
cannot be accounted by this background. Since the neutrino
annihilation takes place in the core of the Sun (which has a size much smaller than
the angular resolution of the detector), it can be considered as a point
source. The optimization of the cuts is done following a blinding
policy. In our case, we optimize the selection cuts, the
search cone size and the quality parameter $\chi^{2}$ in order to
minimize the average upper limit~\cite{hill}:

\begin{equation}
\bar{\phi}_{\nu}^{90\%} = \frac{\bar{\mu}^{90\%}}{A_{\rm eff}(M_{\rm
    WIMP}) \times T_{\rm eff}}, 
\label{phinuaulimiteqn}
\end{equation}

\noindent where $\bar{\mu}^{90\%}$ is the average limit in the number
of events, $A_{{\rm eff}}(M_{\rm WIMP})$ is the effective area and $T_{{\rm eff}}$ is
the active time. The effective area for the cuts  $\chi^2_t < 1.4$ and
$\Psi < 3^{\circ}$ is shown in Fig.~\ref{aeff_channels}. It includes
the Sun visibility (i.e. how much time the Sun is below the horizon)
and the bin efficiency (i.e. how much signal is kept in the search cone).

\section{Results}
\label{results}

\begin{figure}
\includegraphics[width=\linewidth]{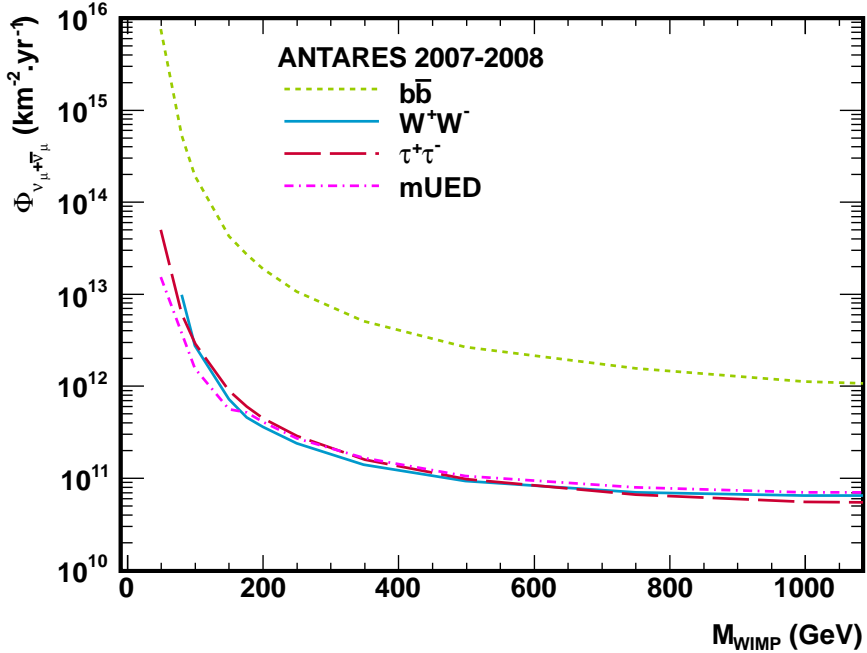}
\caption{Sensitivity in the neutrino flux as a function of the WIMP mass. (Preliminary).}
\label{fluxlimit}
\end{figure}

The average upper limits expected for different models are shown in
Fig.~\ref{fluxlimit} as a function of the WIMP mass. For the case of
CMSSM there is a wide spread in the branching ratios for the different
models inside the framework. Therefore, they are presented separately
for the main channels ($W^+W^-$, $\tau\bar{\tau}$, $b\bar{b}$). The
first two channels offer the best limits since the amount of neutrinos
produced in these channels is higher. The spread for the mUED framework is smaller, so we
present the sensitivity for one particular realization. In this case,
the signal is dominated by the $\tau\bar{\tau}$ channel, which
explains why we obtain similar limits to the corresponding channel in CMSSM.

These limits can be used to constrain the cross section of interaction
between neutralinos or KK particles and protons, in particular for the spin dependent
part. The differential neutrino flux is related to the
annihilation cross section $\Gamma$ by

\begin{equation} 
\frac{d\phi_{\nu}}{dE_{\nu}} = \frac{\Gamma}{4\pi d^{2}} \, \frac{dN_{\nu}}{dE_{\nu}}, 
\end{equation}

\noindent where $d$ is the distance Sun-Earth and $dN_{\nu}/dE_{\nu}$
is the differential number of neutrino events for each channel. If we
assume that equlibrium between capture and annihilation has been
reached in the Sun, we can write (for definition of $C_{\otimes}$ see
Eq.~\ref{captureexp})

\begin{equation}
\Gamma \simeq \frac{C_{\otimes}}{2}.
\label{gammaexp}
\end{equation}

On the other hand, the capture rate is related to the spin-dependent scattering cross-section
between a WIMP and a proton as

\begin{multline}
C_{\otimes} \simeq 3.35 \times 10^{18} {\rm s^{-1}} \times
\left(\frac{\rho_{\rm local}}{0.3 \, GeV \cdot {\mathrm cm^{-3}}}\right) \times \\
\times \left(\frac{270 \, {\mathrm km}\cdot {\rm s^{-1}}}{v_{\rm local}}\right) 
\times \left(\frac{\sigma_{H,SD}}{10^{-6} \, {\rm pb}}\right) \times
\left(\frac{TeV}{ M_{\rm WIMP}}\right)^{2}, 
\label{captureexp}
\end{multline}


\noindent where $\rho_{\rm local} = 0.3$ GeV$\cdot$cm$^{-3}$ is the
local density of WIMPs, $v_{\rm local} = 270$ km$\cdot$s$^{-1}$ is the
local mean velocity of WIMPs assuming a Maxwell-Boltzmann velocity
distribution, \\$M_{\rm WIMP}$ is the mass of the dark matter particle
and $\sigma_{H,SD}$ is the spin-dependent scattering cross-section
between a WIMP and a proton.

We have also used the package SuperBayes~\cite{superbayes} and a modified
version of it~\cite{superbayes2} in order to scan the available parameter space of the
CMSSM and mUED models, which are also plotted. It can be seen that neutrino
telescopes like ANTARES offer the best limits for spin-dependent cross-section.

\begin{figure}[c]
\begin{center}
\includegraphics[width=\linewidth]{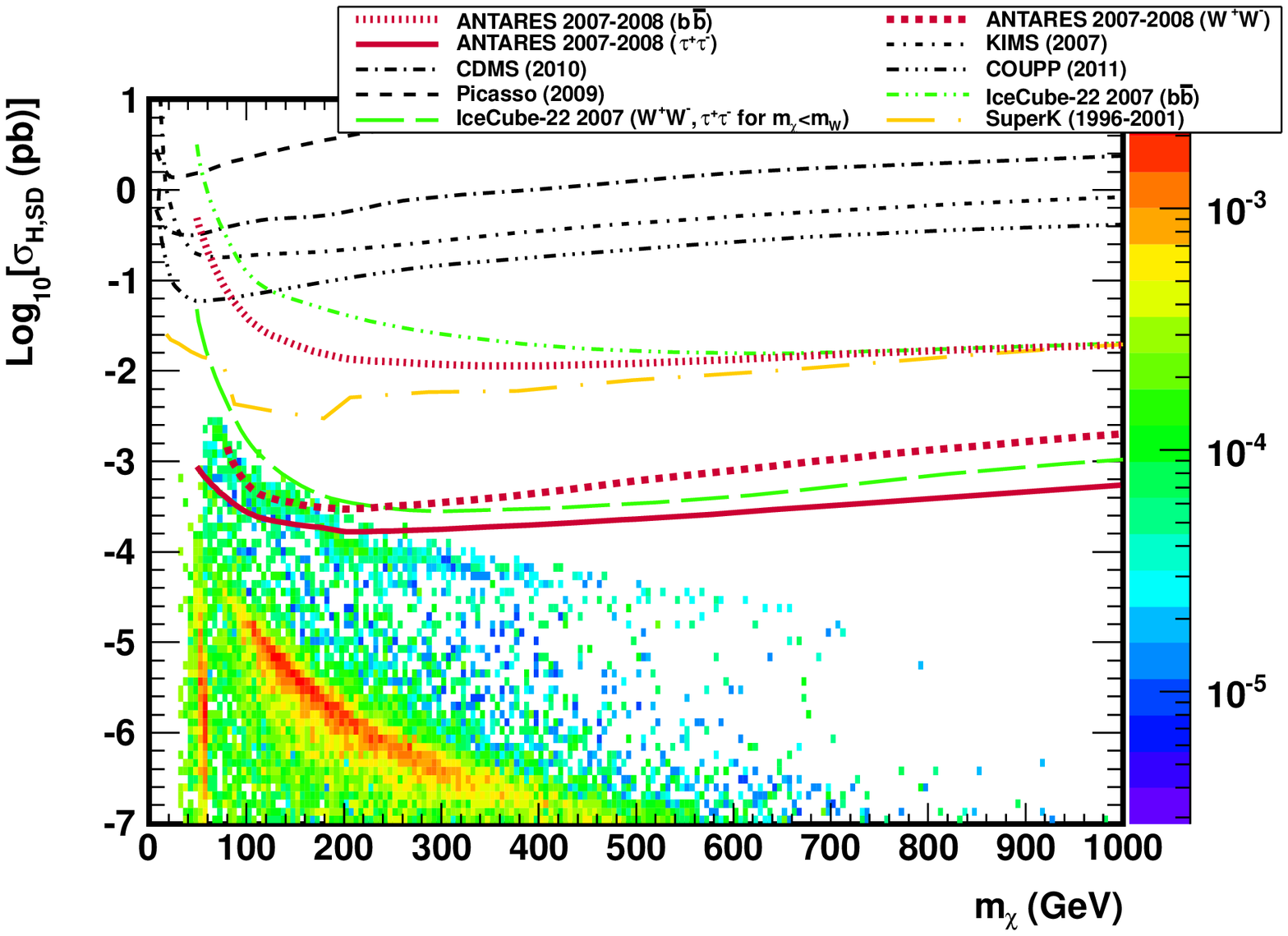}  \\
\includegraphics[width=\linewidth]{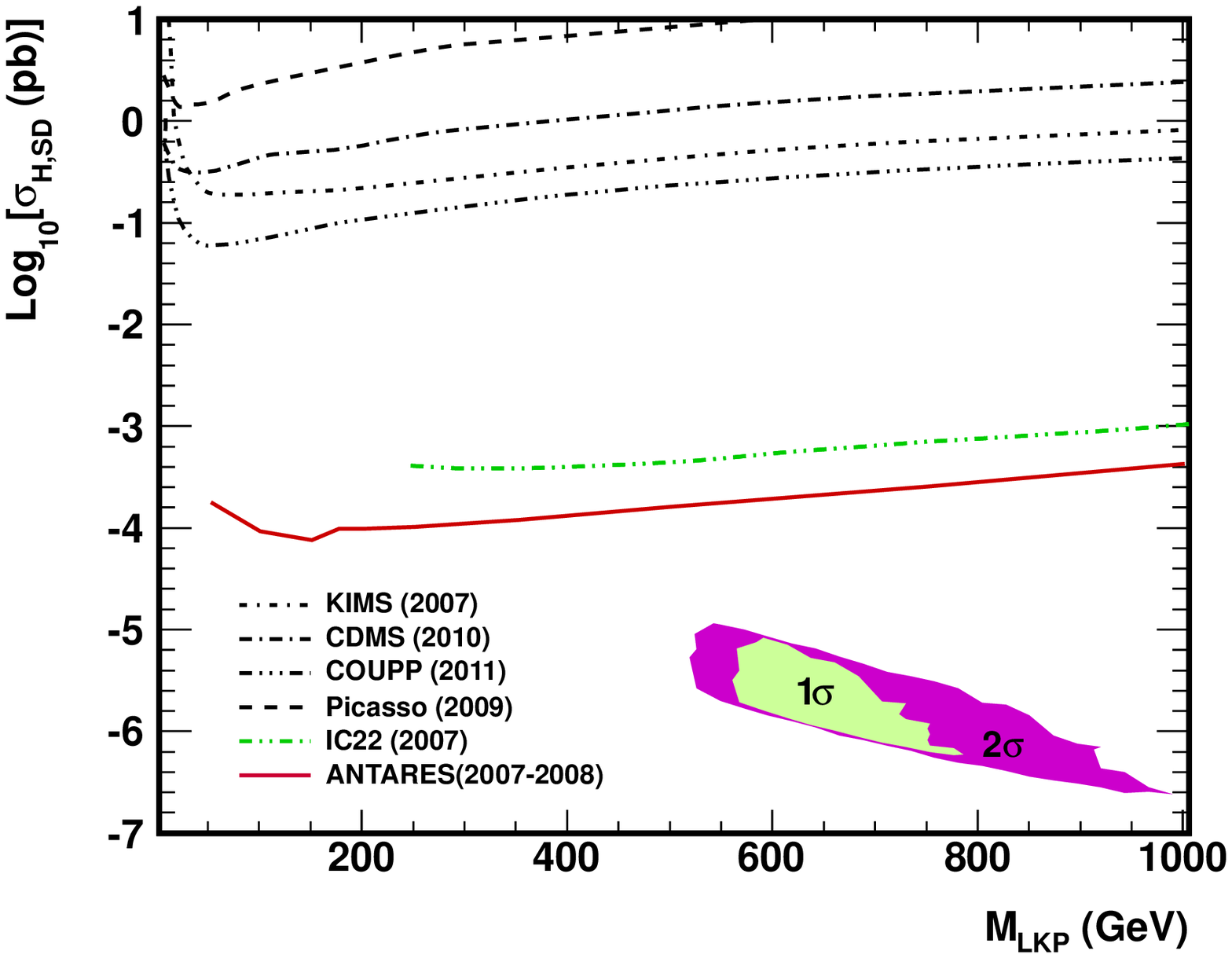}
\caption{Experimental limits in the spin-dependent cross-section
  $\bar{\sigma}_{H,SD}$ compared to the theoretical parameter space allowed
  by the experimental constraints. The upper plot corresponds to the
  CMSSM model. The bottom plot corresponds to the mUED model, where
  the light and dark areas correspond to the one and two sigma exclusion
  regions, respectively. (Preliminary).}
\label{final}
\end{center}
\end{figure}

\section{Conclusions}
\label{conclusions}

We have presented the sensitivity of the search for neutralinos in the
Sun with the ANTARES telescope. We have used data of 2007 and 2008,
when the detector configuration had 5 lines (2007) and 10, 9 or 12
(2008). We have studied two frameworks, CMSSM and mUED, and the
average limits on the neutrino flux and the expected limits for the
spin dependent cross section of the interaction of WIMPs with
protons.

\section*{Acknowledgements}
The author acknowledges the support of the Spanish MICINN’s
Consolider-Ingenio 2010 Programme under grant MultiDark CSD2009-00064.
\label{acknow}


\end{document}